\documentclass[aps,prl,twocolumn]{revtex4}
\usepackage{amssymb}
\usepackage{natbib}
\usepackage{amsmath}
\usepackage{amssymb}
\usepackage{amsfonts}
\usepackage{graphicx}
\usepackage{mathrsfs}
\usepackage{dcolumn}
\usepackage{bm}
\usepackage{color}

\definecolor{rot}{rgb}{0.75,0.05,0.25}
\definecolor{hellgrau}{gray}{0.5}
\definecolor{blau}{rgb}{0,0,0.7}

\begin{document}

\title{Reply to ``Comment on ``On a new definition of quantum entropy''''}
\author{Michele Campisi}
\email{campisi@unt.edu} \affiliation{Department of
Physics,University of North Texas Denton, TX 76203-1427, U.S.A.}
\date{\today }

\begin{abstract}
The example provided in the comment [arXiv:0803.2241] concerns a
situation where the system is initially at negative temperature.
It is known that in such cases the Law of Entropy Decrease holds.
Nevertheless, this does not challenge the validity of the Second
Law of Thermodynamics.
\end{abstract} 
\maketitle

In the comment \cite{Sadri08} on ``On a new Definition of Quantum
entropy'' \cite{Campisi08bis}, the author argues against the
convenience of the new entropy operator
\begin{equation}\label{eq:S=logN}
    \hat{\mathcal{S}}(t) \doteq \ln (\hat{\mathcal{N}}(t)+\hat{1}/2)
\end{equation}
on the basis of the fact that for an ensemble of identical quantum
systems with finite spectrum, for which only the highest level is
populated, one observes a decrease of the expectation of the
entropy (\ref{eq:S=logN}) and not an increase.

In regards to this we would like to point out that (see
\cite{Campisi08}) the expectation value of the entropy in
(\ref{eq:S=logN})
\begin{itemize}
    \item[a)] \emph{increases} whenever the initial
probability distribution is \emph{decreasing}
    \item[b)] \emph{decreases} whenever the initial
probability distribution is \emph{increasing}
\end{itemize}
A decreasing probability $p_n$ is characterized by the condition
\begin{equation}\label{eq:pm>pn}
    p_m \leq p_n \quad \text{if} \quad m > n
\end{equation}
An increasing probability $p_n$ is characterized by the condition
\begin{equation}\label{eq:pm<pn}
    p_m \geq p_n \quad \text{if} \quad m > n
\end{equation}
The latter is often referred to as an ``inverted population'',
which characterizes systems at ``negative temperature''.

It is evident that the example provided in the comment
\cite{Sadri08} belongs to the case of negative temperature, and
this is the reason why one observes a decrease of entropy.

It has been discussed in \cite{Campisi08} that the existence of a
Law of Entropy Decrease in systems at negative temperature does
not challenge the validity of the Second Law of Thermodynamics.
This is because the \emph{natural} state of matter is at positive
temperature (Gibbs state), whereas negative temperatures can only
be created artificially by spending entropy and work. So the
\emph{natural} evolution of the quantum entropy is towards larger
values although it may move to lower values in \emph{artificially}
created situations.

The example of laser functioning well explains this point. To
operate a laser we first create an inverted population from a
Gibbs state at positive temperature that Nature provides us ``for
free''. Secondly we make the excited states decay thus emitting a
radiation. During the first step we spend entropy (the entropy of
the system increases), whereas during the second step we gain
entropy (the entropy of the system decreases). If we consider the
total process consisting of the two steps, we immediately see that
it is characterized by an overall \emph{increase of entropy}
because the very \emph{initial state was at positive temperature}.

It must be stressed that, when operating a laser, an analogous
\emph{work balance} occurs \cite{Allahverdyan02}. In particular
the laser functioning is based on the cyclic extraction of work
from the negative temperature system during the second step. This
is seemingly in contradiction with the second law as expressed by
Thomson, according to which no work extraction is possible by
means of cyclic transformation \cite{Allahverdyan02}. The
contradiction is resolved once the work spent to create the
inverted population is properly counted in the work balance.

Thus, in contrast to the intuitive expectation both Thomson and
Clausius formulations \emph{are indeed inverted} when an ensemble
of excited states decays (if this was not true then we could not
use lasers to our own advantage!). Nonetheless the overall balance
of work and entropy conforms to the Second Law. Therefore, neither
the convenience of work definition used in \cite{Allahverdyan02},
nor the convenience of the new definition of entropy in Eq.
(\ref{eq:S=logN}), are to be questioned on these grounds. On the
contrary they help us understand the quantum origins of the Second
Law, and provide a coherent Quantum Thermodynamic scheme for
studying the efficiency of both positive and negative temperature
devices.

A more detailed discussion of the properties and fundamental
implications of the new entropy definition can be found in
\cite{Campisi08}.

\end{document}